\documentclass[]{aastex}
\usepackage{emulateapj5}

\shorttitle{Formation of exponential disks from clump clusters}

\shortauthors{Bournaud, Elmegreen \& Elmegreen}

\begin{document}

\title{Rapid formation of exponential disks and bulges at high redshift\\
from the dynamical evolution of clump cluster and chain galaxies}

\author{Fr\'ed\'eric Bournaud \affil{Laboratoire AIM, CEA/DSM - CNRS - Universit\'e Paris Diderot\\Dapnia/Service d'Astrophysique,
CEA/Saclay, F-91191 Gif-sur-Yvette Cedex, France,
frederic.bournaud@cea.fr} }
\author{Bruce G. Elmegreen \affil{IBM Research Division, T.J. Watson
Research Center, P.O. Box 218, Yorktown Heights, NY 10598, USA,
bge@watson.ibm.com} }
\author{Debra Meloy Elmegreen \affil{Vassar College,
Dept. of Physics \& Astronomy, Box 745, Poughkeepsie, NY 12604;
elmegreen@vassar.edu} }

\begin{abstract}
Many galaxies at high redshift have peculiar morphologies dominated
by $10^8-10^9$ M$_\odot$ kpc-sized clumps. Using numerical
simulations, we show that these ``clump clusters'' can result from
fragmentation in gravitationally unstable primordial disks. They
appear as ``chain galaxies'' when observed edge-on. In less than
1~Gyr, clump formation, migration, disruption, and interaction with the disk cause
these systems to evolve from initially uniform disks into regular
spiral galaxies with an exponential or double-exponential disk profile and a
central bulge. The inner exponential is the initial disk size and
the outer exponential is from material flung out by spiral arms and
clump torques. A nuclear black hole may form at the same time as the
bulge from smaller black holes that grow inside the dense cores of
each clump. The properties and lifetimes of the clumps in our models
are consistent with observations of the clumps in high redshift
galaxies, and the stellar motions in our models are consistent with
the observed velocity dispersions and lack of organized rotation in
chain galaxies. We suggest that violently unstable disks are the
first step in spiral galaxy formation. The associated starburst
activity gives a short timescale for the initial stellar disk to form.
\end{abstract}


\section{Introduction}

\begin{figure}
\centering
\includegraphics[width=4.5in]{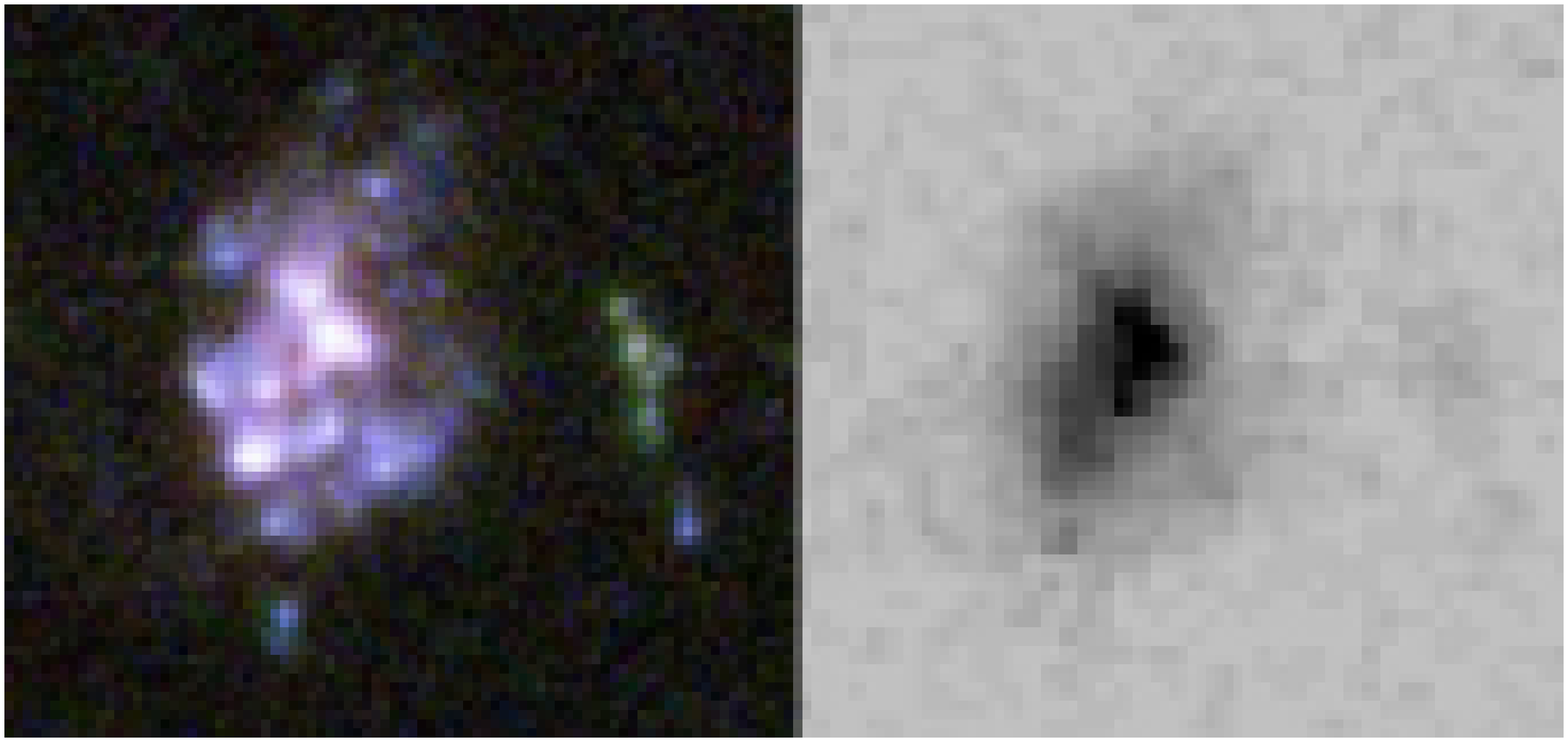}\vspace{.2cm}\\
\includegraphics[width=4.5in]{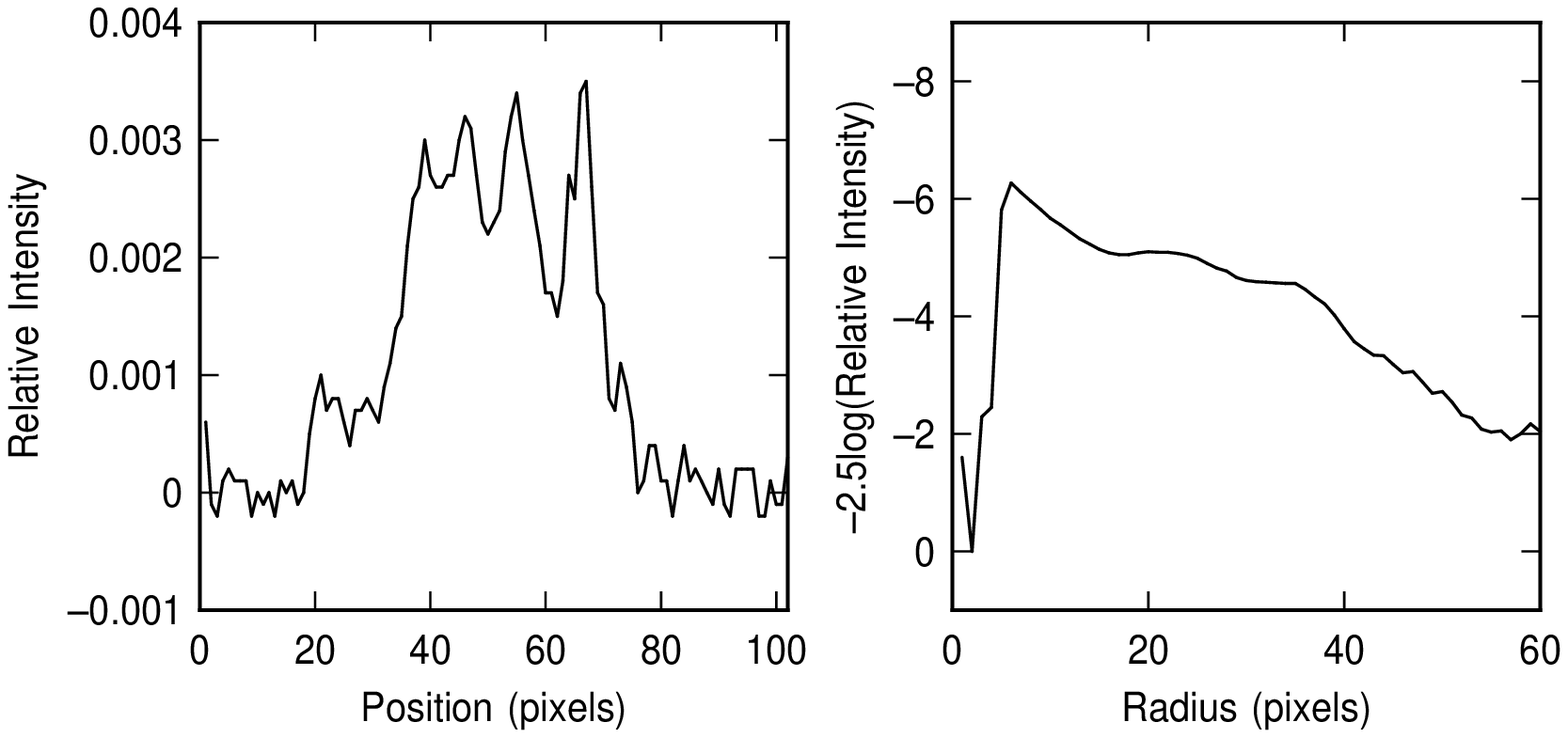}
\caption{upper: Clump-cluster galaxy UDF 1666 in a color optical
image from the UDF skywalker (left) and in NICMOS H band ( right).
lower: (left) The luminosity profile is shown along the major axis,
through the apparent center of the galaxy; this profile is irregular
and rather flat. (right) The azimuthally averaged radial profile has
a dip in the center, a peak at the brightest spot, and declines
gradually outward as the number density of clumps decreases.  The
bumps correspond to radii where the main clumps occur, inside about
40 pixels. The position scale in pixels corresponds to 0.03 arcsec
per pixel, or about 250 pc at the galaxy's photometric redshift of 1.4. This
profile and the corresponding image resemble step 2 in Figures 6 and
4.}\label{fig:udf1666}\end{figure}

Galaxies at redshifts larger than $z\sim1$ become increasingly
clumpy with star formation in kpc-size complexes containing
$10^7-10^9$M$_\odot$ that are several hundred Myr old (see review in
Elmegreen 2007). A remarkable morphology is that of ``chain
galaxies'' with large visible clumps aligned along one axis (Cowie,
Hu \& Songaila 1995; van den Bergh et al. 1996; Moustakas et al.
2004). These are rare in the Local Universe (e.g., Abraham et al.
1996) but much more frequent at redshift $\sim 1$ (Elmegreen et al.
2005a, hereafter EERS05). Dalcanton \& Schectman (1996) suggested that chain galaxies
could be edge-on LSBs. However, the clumps are much more massive
than those observed in the UV in nearby edge-on disks (e.g., Smith
et al. 2001), and the actual face-on counterparts of chain galaxies
at high-redshift are observed to be starburst disks with
kpc-sized clumps -- the so-called ``clump cluster galaxies''
(Elmegreen, Elmegreen \& Hirst 2004, see the example of
UDF~1666 in Fig.~\ref{fig:udf1666} from EERS05). Both clumpy types are found up
to the bandshifting limit of $z\sim5$ (Elmegreen et al. 2007a,
hereafter EERC07). Ring (Elmegreen \& Elmegreen 2006a) and
interacting (Elmegreen et al. 2007b) galaxies at high redshift are
also very clumpy. Distant ellipticals can be clumpy too, as found in
the Tadpole and A1689 cluster fields (Menanteau et al. 2004, 2005)
and the Ultra Deep Field (UDF; Elmegreen, Elmegreen, \& Ferguson
2005). Clumpiness in high redshift galaxies has been quantified with
the $S$ parameter by Conselice (2003) and Conselice et al. (2004).

O'Neil et al. (2000) suggested that chains could be knotty disks
seen edge-on; otherwise the edge-on counterparts of face-on disks
would be missing at high redshift. Taniguchi \& Shioya (2001)
proposed that chains are filaments of clumps about to merge into
elliptical remnants. Detailed observations actually confirm that
{\it chain} and {\it clump-cluster} galaxies are mostly a single
class of objects that are clumpy disks viewed in different
orientations: the two types have equivalent sizes, magnitudes, and
redshift ranges, and their clumps have similar properties (Elmegreen
\& Elmegreen 2005, hereafter EE05, and EERC07). Their combined
distribution of axial ratios is flat, as it is for a single disk
population (EE05). In a sample of 10 extremely clumpy galaxies in
the UDF, the fraction of the light in the clumps was found to be
$\sim40$\% and the fraction of the total stellar mass was $\sim10$\%
(EE05). On average in the UDF, the luminosity fraction of the clumps
in 178 clump-cluster galaxies is $\sim 20$\%, and the luminosity
fraction of the clumps in 269 spiral galaxies is $\sim5$\%, not
counting the bulges in the latter (EERS05).
Comparisons between ACS and NICMOS images of clump cluster and chain
galaxies confirm that the clumps are intrinsic to the mass and are
not rest-frame blue patches on a smooth underlying disk (EERC07).
The clumps therefore represent a large fraction of the mass in these
galaxies, unlike the clumps in modern spirals.

Clump clusters and chains have irregular and somewhat flat
luminosity profiles -- different from spirals, which have bulges and
exponential disks.  However, the azimuthally averaged number density
of clumps as a function of radius in UDF clump clusters, when
normalized to the galaxy sizes, is similar to that of the (smaller)
clumps in the spirals, and both distributions are close to
exponential (Elmegreen et al. 2005b, hereafter EEVFF05). This
equivalence suggests that exponential disks and bulges in spiral
galaxies generally form by the dissolution of clumps in clump
cluster and chain galaxies. Consistent with this scenario is the
observation that the earliest disks in the Universe are clump
clusters and chains (EERC07).  Thus, clump cluster galaxies may
evolve into spirals, making the ``proto-spiral'' nomenclature for
this type in van den Bergh et al. (1996) remarkably prescient.

The purpose of this article is to simulate the evolution of
clump-cluster galaxies and determine their final state. Noguchi
(1999) and Immeli et al. (2004a,b) also simulated gravitational
collapse in highly gaseous disks and found that they formed giant
clumps that interact and eventually merge into a bulge. They did not
consider the final disk profile and their galaxies had the masses
and sizes of modern galaxies. Here we consider disks that are about
half of this size as observed in the UDF (EERC07). We show that
initially uniform unstable disks evolve quickly through a clump
cluster phase and end up with bulges and double-exponential
profiles. The absence of clear rotation in chain and clumpy galaxies
(Bunker et al. 2000; Erb et al. 2004; Weiner et al. 2006) is also
explained: the massive clumps severely affect the velocity fields
and enhance the disk velocity dispersion.

The numerical techniques and initial conditions are described in
Sect.~2. We study the properties and evolution of clump cluster and
chain galaxies and the formation of exponential disks in Sect.~3. In
Sect.~4, we discuss the results and compare with the evolution of
stable spiral disks. Sect.~5 considers the formation of nuclear
black holes. Our conclusions are summarized in Sect.~6.

\section{Simulations}

\subsection{Code description}

We model the evolution of Jeans-unstable gas-rich galactic disks
using a particle-mesh sticky-particle code (see Bournaud \& Combes
2002, 2003). The grid resolution and gravitational softening are
110~pc. Stars, gas, and dark matter halo are modeled with one
million particles each. Unless specified otherwise, we use
sticky-particle parameters $\beta_r=\beta_t=0.7$ (defined as in
Bournaud \& Combes 2002).

Star formation is described by a local Schmidt law. At each
timestep, the probability for each gas particle to be transformed
into a stellar particle is proportional to the local gas density to
the exponent 1.4 (e.g., Kennicutt 1998), the proportionality factor
being computed to provide a star formation rate of
$3.5$~M$_{\sun}$~yr$^{-1}$ in the initial uniform disk, equivalent
to a timescale for gas consumption of 5~Gyr (before the formation of
dense clumps increases the star formation rate). Feedback from star
formation is not implemented in this study; the consequences of this
choice are discussed in Section~4.2.

\subsection{Initial conditions}

We start the simulations with a flat and bulgeless disk of gas and
stars, which is Jeans-unstable and has a uniform surface density.
This is because clump cluster galaxies are not observed to have a
bulge, and they have irregular radial profiles. For instance, 
the typical clump cluster galaxies UDF~3752
and UDF~82+99+103 in EEVFF05, and UDF~1666 shown here in
Figure~\ref{fig:udf1666}, have an irregular profile, which on large
scales is rather flat.
The azimuthally averaged radial profiles for these three galaxies
are irregular as well, showing a central minimum for UDF~82+99+103
and UDF~1666, prominent bumps at several radii, and either a
flattening (UDF~3752) or a slow taper from the gradient in clump
number density out to the edge (UDF~1666). 
To study the formation of exponential disks from clump clusters we
thus have to start with a disk that does not already have a
concentrated nearly-exponential profile. The instabilities will
rapidly make it irregular.

\begin{figure}
\centering
\includegraphics[width=3in]{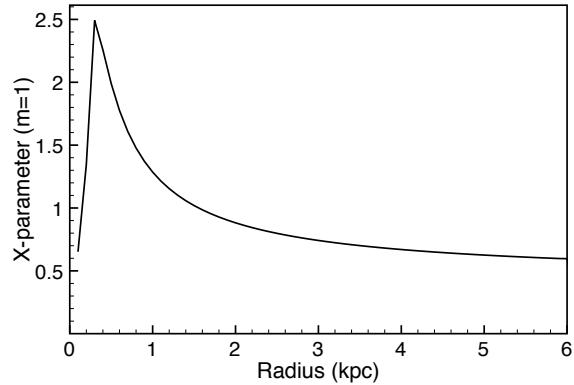}
\caption{X-parameter (Toomre 1964) in the initial disk for the fiducial run~0, in the $m=2$ mode. Because of the initial flat profile of the disk and absence of bulge, it differs from typical spirals with values $\sim 1$ except in the central region.}\label{fig:xparam}\end{figure}

The initial disk radius is 6~kpc and its mass $7 \times
10^{10}$~M$_{\sun}$, comparable to the observed properties of $z
\simeq 1$ systems (e.g., EEVFF05). The initial thickness is
$h=700$~pc with a $\mathrm{sech}^2\left( z/h \right)$ vertical mass
distribution, which ensures that, once instabilities begin to heat
the system, the thickness is compatible with the observed average of
$h=1$~kpc (fitted on $\mathrm{sech}^2\left( z/h \right)$ profiles,
Elmegreen \& Elmegreen 2006b). Stars initially have a Toomre
parameter $Q_{\mathrm{s}}=1.5$. The dark halo is a Plummer sphere of
radial scale-length 15~kpc. The velocity dispersion of the gas
$\sigma$, gas mass fraction in the disk $f_{\mathrm{G}}$, and
halo-to-disk mass ratio $H/D$ (measured inside the initial disk
radius) are varied as indicated in Table~1. The fiducial run~0 has
$\sigma=9$~km~s$^{-1}$, $f_{\mathrm{G}}=0.5$ and $H/D=0.5$, the
resulting Toomre X-parameter (Toomre 1964) is shown as a function of
radius on Figure~\ref{fig:xparam}; this parameter is defined by:
\begin{equation}
X(r)=\frac{r \kappa(r)^2}{2 \pi G m \mu(r)}
\end{equation}
where $\kappa$ and $\mu$ are the epicyclic frequency and surface density, respectively. We show on Figure~\ref{fig:xparam} the value of $X$ for an azimuthal wavenumber $m=2$.

For further comparisons, we also performed a control run (run C-4)
with the same mass distribution as run~4, but initially stable with
$Q_{\mathrm{s}}=1.8$ and a gas turbulent speed computed at each
radius to ensure $Q_{\mathrm{g}}=1.1$. The sticky-particle
parameters are changed to $\beta_r=\beta_t=0.8$ to avoid global disk
instabilities from kinematical cooling. As a result of these
changes, C-4 follows the evolution of a spiral galaxy instead of a
clump cluster. We ran this test analogous to run~4 because this run
has the lowest gas fraction and is the most easily stabilized with a
modest gas turbulent speed.

We also changed the sticky-particles parameters in run SP-0: the
initial conditions are that of run~0, but $\beta_r=0.5$ and
$\beta_t=1$ to conserve angular momentum with about the same rate of
energy dissipation.

\subsection{Clump detection and analysis}

Because the simulated disks are unstable to fragmentation, we need
to detect the large clumps that form in order to measure their
masses and other related quantities. Because observations define
clumps as morphological entities, we chose to detect clumps in the
simulated galaxies from morphological criteria too. This is made
every 50~Myr, on face-on projected density maps, smoothed at a
resolution of 200~pc to decrease the particle noise. We compute the
average surface density as a function of radius $\mu_0 (r)$. First,
clumps must represent local overdensities compared to the average
density at the same radius. We then keep the pixels with $\mu > 3
\mu_0 (r)$  \footnote{This choice was made because in some cases
overdensities with $\mu > 2 \mu_0 (r)$ were found to include spiral
arms in spite of the maximum size criterium.}. We consider only the
regions less extended than 3~kpc, which was found to eliminate
spiral arms. The remaining regions are considered as ``clumps'' if
their mass is at least $2\times10^8$~M$_\sun$, for we consider only
the high-mass clumps in this study. The regions obtained this way
are split into two clumps if there are two luminosity peaks $\mu_1$
and $\mu_2$ that are separated by pixels with $\mu< 1/3 \min
(\mu_1,\mu_2)$, i.e. a significant interclump contrast, and each new
clump is above the mass threshold (we do not separate low-mass
subclumps). An example of clump detection is shown on
Figure~\ref{fig:detect}.

To better understand the role of clumps in fueling the bulge and the
disk (see below), we measured several mass fractions that are
indicated for each run in Table~1. All these fractions are defined
within the baryonic mass (gas and stars), not counting the dark
matter halo component.
\begin{itemize}
\item $f_\mathrm{C}$ is the fraction of the baryonic mass that is
concentrated in the clumps, taken at the instant of its highest
value, the clump borders being defined as explained above.
\item $f_\mathrm{g,C}$ is the gas fraction in the clumps, which is the fraction of the mass in the clumps defined above that is in the form of gas. The value in Table~1 is given at the same instant as $f_\mathrm{C}$.
\item $f_\mathrm{B,C}$ is the fraction of the mass from the clumps (counted in $f_\mathrm{C}$) that is in the bulge at the end of the simulation. $1-f_\mathrm{B,C}$ is the mass fraction of the clumps that ends up in the disk.
\item $f_\mathrm{C/B}$ is the fraction of the bulge mass that comes from the clumps, $1-f_\mathrm{C/B}$ being the fraction of the bulge that does not come from the clump material.
\item $f_\mathrm{C/D}$ is the fraction of the disk mass that comes from the
clumps.
\end{itemize}

In these definitions, the {\it disk} mass is obtained by the
integration of the fitted double-exponential profile described
below. The {\it bulge} mass is then defined as the mass left over
above this disk profile in the central regions, without any
consideration of the bulge luminosity profile itself. Bulges were
never found to contribute to the luminosity profile  beyond
$r=2$~kpc, so we decide that all particles at radii larger than
2~kpc are disk particles. Particles at radii $r<2$~kpc have a
probability $\mu_{\mathrm{d}}(r)/\mu(r)$ to be disk particles, where
$\mu(r)$ is the measured (total) surface density profile and
$\mu_{\mathrm{d}}$ the (fitted) density profile of the disk
component. The probability that it is a bulge particle is
$1-\mu_{\mathrm{d}}(r)/\mu(r)$. These two probabilities are limited
to 1 and 0 respectively when fluctuations cause $\mu(r)$ to be
larger than $\mu_{\mathrm{d}}(r)$. Thus, the definition of bulge and
disk particles in the central regions is made on a simple
statistical basis that retrieves the mass and radial distribution of
each component. More advanced criteria might distinguish the bulge from the disk
based on vertical distributions and velocity dispersions. However,
significant differences from the present definitions would appear
only in a narrow radial region around $r\simeq1$ kpc, because at other
radii one or the other of the two components dominates.

\section{Results}

\subsection{Clump cluster evolution, disk and bulge fueling}
\label{sect:evol}

\begin{figure}
\centering
\includegraphics[width=3.9in]{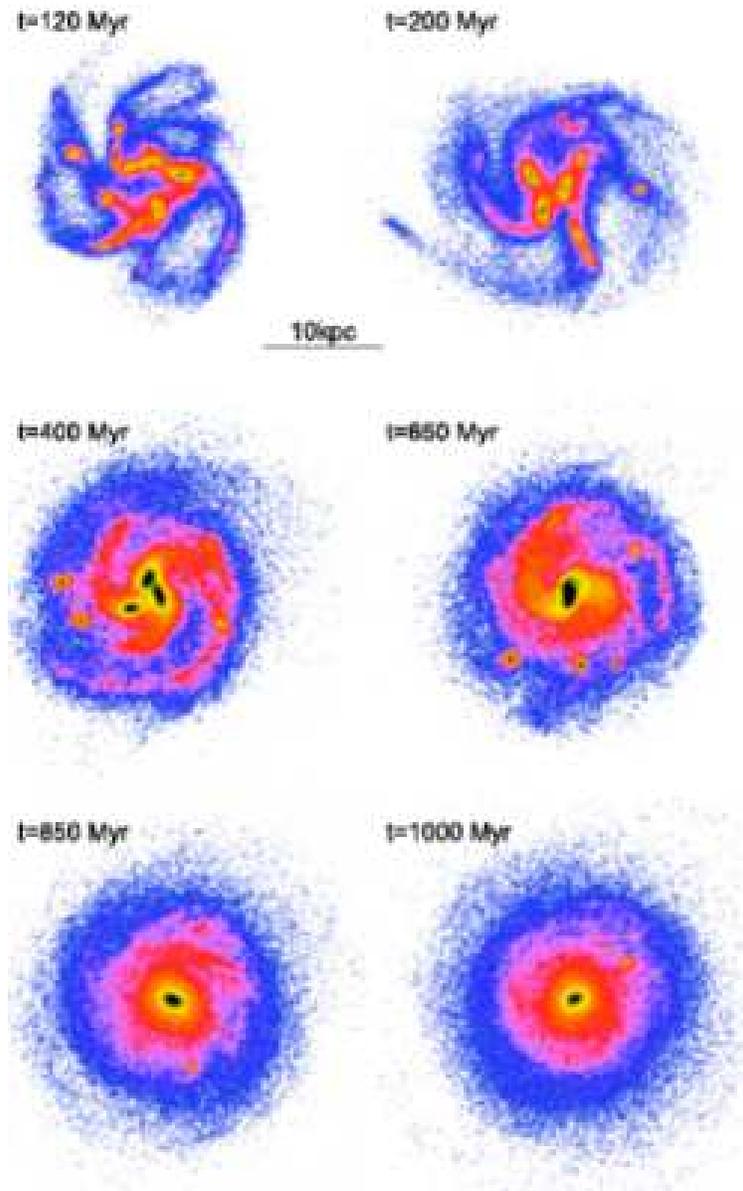}
\caption{Face-on snapshots of the mass surface density in the fiducial run~0, showing the formation of a clump cluster galaxy and its evolution towards a classical spiral galaxy with a central bulge and (double-) exponential disk.}\label{fig:evol}\end{figure}

Figure~\ref{fig:evol} shows the evolutionary sequence for the
fiducial run. The initial bulgeless disk is gravitationally unstable
and makes several large clumps of stars and gas quickly. The clumps
last $\simeq 400$ Myr and gather $f_\mathrm{C}=38\%$ of the total
baryonic mass of the system at $t=200$ Myr, which is the time at
which this fraction reaches its maximum. The mass fraction of the
stars in these clumps compared to the total stellar mass is 29\%. At
this stage the clumps are gas-rich with a gas mass fraction of 56\%,
while the gas fraction over the whole galaxy at the same instant is
42\%.

\begin{figure}
\centering
\includegraphics[width=3.3in]{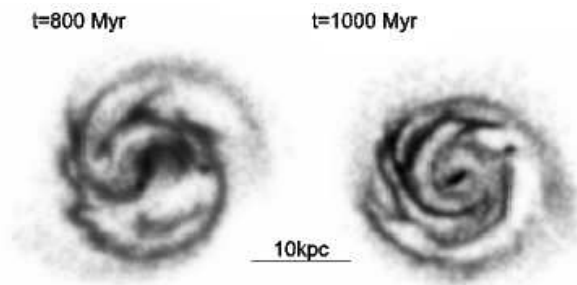}
\caption{Face-on snapshot of the gas mass density in the fiducial run~0 (total mass density is shown on Figure~\ref{fig:evol}) at two late instants, showing the spiral structure of the final galaxy after the clump-cluster phase.}\label{fig:gas}\end{figure}

\begin{figure}
\centering
\includegraphics[angle=0,width=3in]{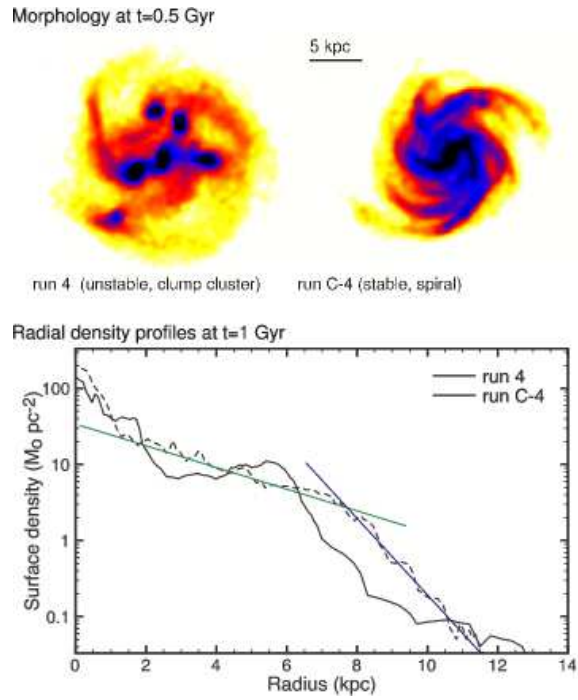}
\caption{Radial density profiles at $t=1$~Gyr for the run~4
(clump-cluster evolution) and run C-4 (stable spiral disk with the
same initial mass distribution). The clump-cluster redistributes the
initial flat disk into an exponential-like disk more efficiently
than density waves within this timescale. The snapshots of these two
runs are seen at $t=0.5$~Gyr, when the clumps are still present for
the unstable model.}\label{fig:control}\end{figure}

The clumps then progressively release their mass in the disk: at
$t=400$~Myr the clump mass fraction has decreased to 27\% from the
initial 38\%.
They have formed stars rapidly with an average star formation rate
between $t=200$ and $t=400$~Myr of 38~M$_{\sun}$~yr$^{-1}$, and the
gas mass fraction in the clumps has decreased from 56\% to 37\%. The
star formation rate in the model is compatible with the observed
average of 20~M$_{\sun}$~yr$^{-1}$ in the real clumps (EE05),
although the star formation efficiencies cannot be directly compared
because of the unknown gas mass.  At the same time, their mutual
interaction and the friction on the halo and underlying disk bring
the clumps to the galaxy center (Fig.~\ref{fig:evol}) within a
timescale of about 4 rotations, which is 0.5--0.7 Gyr. Later on, the
galaxy resembles a spiral galaxy, with a central bulge and only a
little mass in small remaining clumps. The final system is dominated
by stars and shows weak density waves in the mass distribution
(Fig.~\ref{fig:evol}), which can be compared to smooth spiral arms
in the old stellar populations of a spiral galaxy seen in
near-infrared. The gas distribution at two late stages is shown in
Figure~\ref{fig:gas} and has a stronger spiral structure. Other
examples of simulated clump clusters are for run~6 in
Figure~\ref{fig:rc} and run~4 in Figure~\ref{fig:control}.

The models in Noguchi (1999) and Immeli et al. (2004a,b) started
with more massive disks and developed larger clumps, most of which
migrated to the central spheroid, leaving only a faint disk whose
radial profile was not considered. With our initial conditions more
representative of high-redshift clump clusters, we obtain lower
bulge mass fractions: the bulge-to-total mass ratio (including the
halo) varies between 0.12 and 0.36 over our sample of simulations
(Table~1), which correspond to bulge-to-disk ratios between 0.14 and
0.50. Hence, the galaxies resulting from clump cluster evolution are
still disk-dominated galaxies after the bulge has formed.

\subsection{Mass distribution and radial profile}

\begin{figure}
\centering
\includegraphics[angle=0,width=2in]{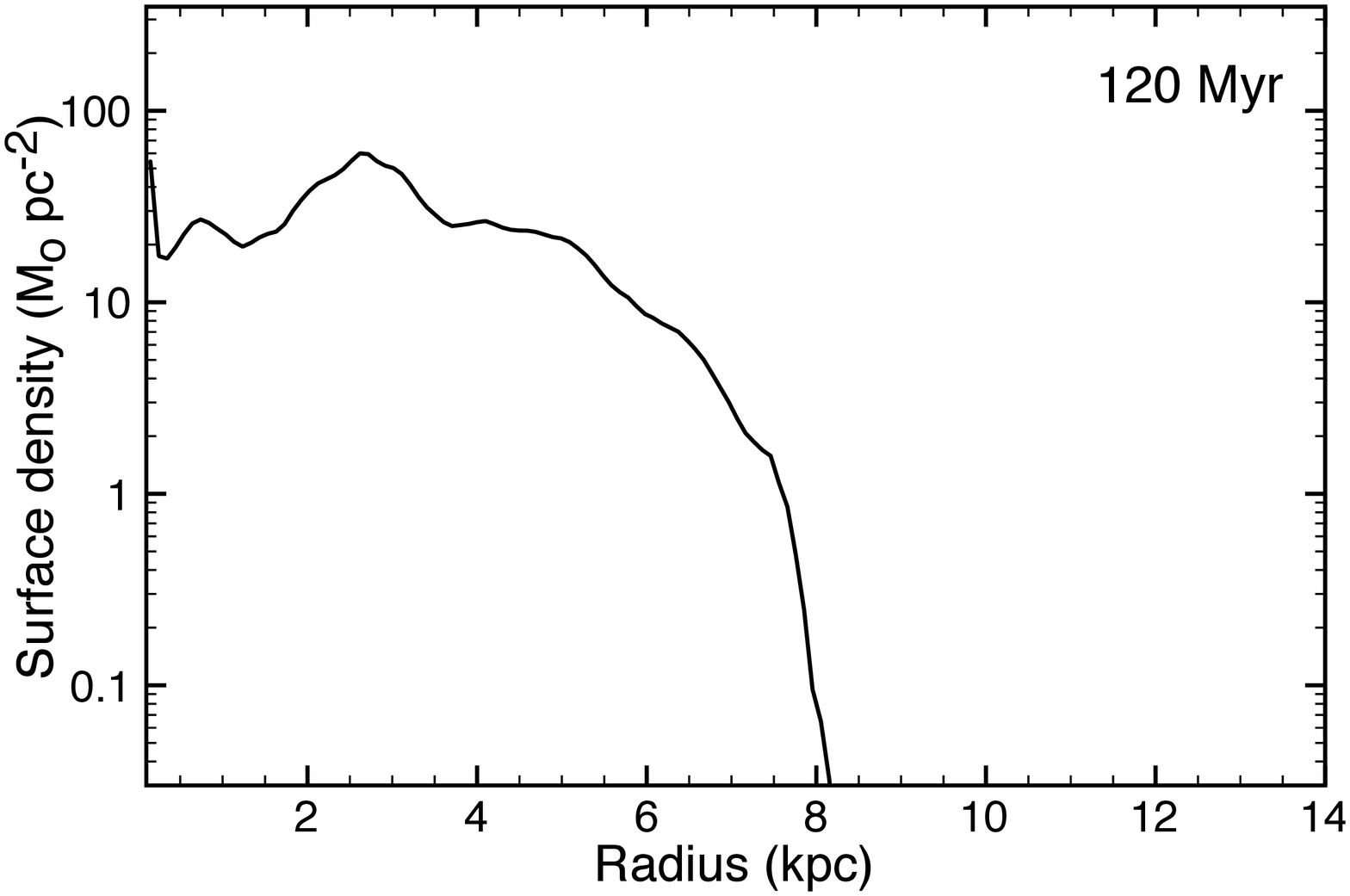}
\includegraphics[angle=0,width=2in]{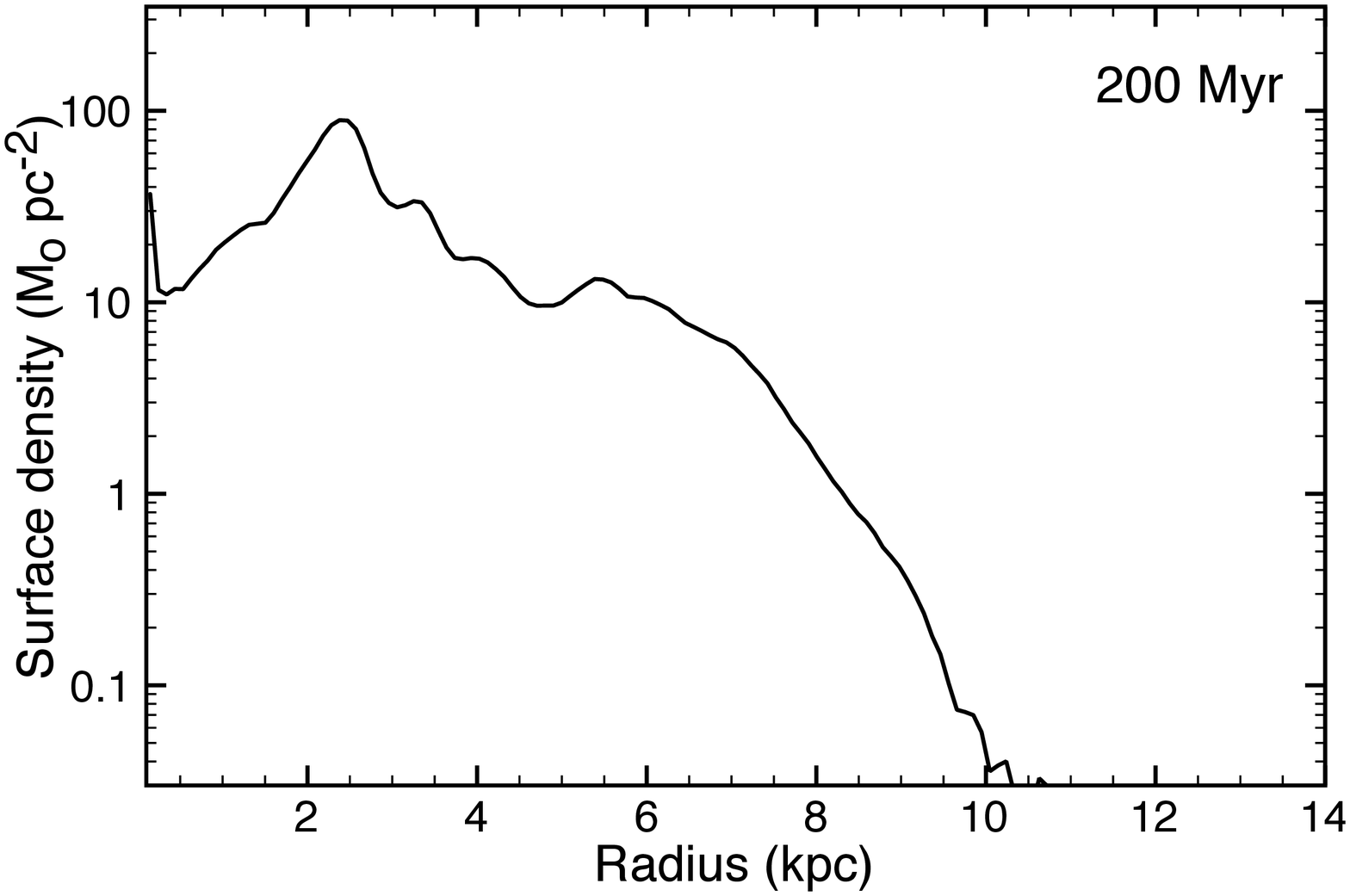}\\
\includegraphics[angle=0,width=2in]{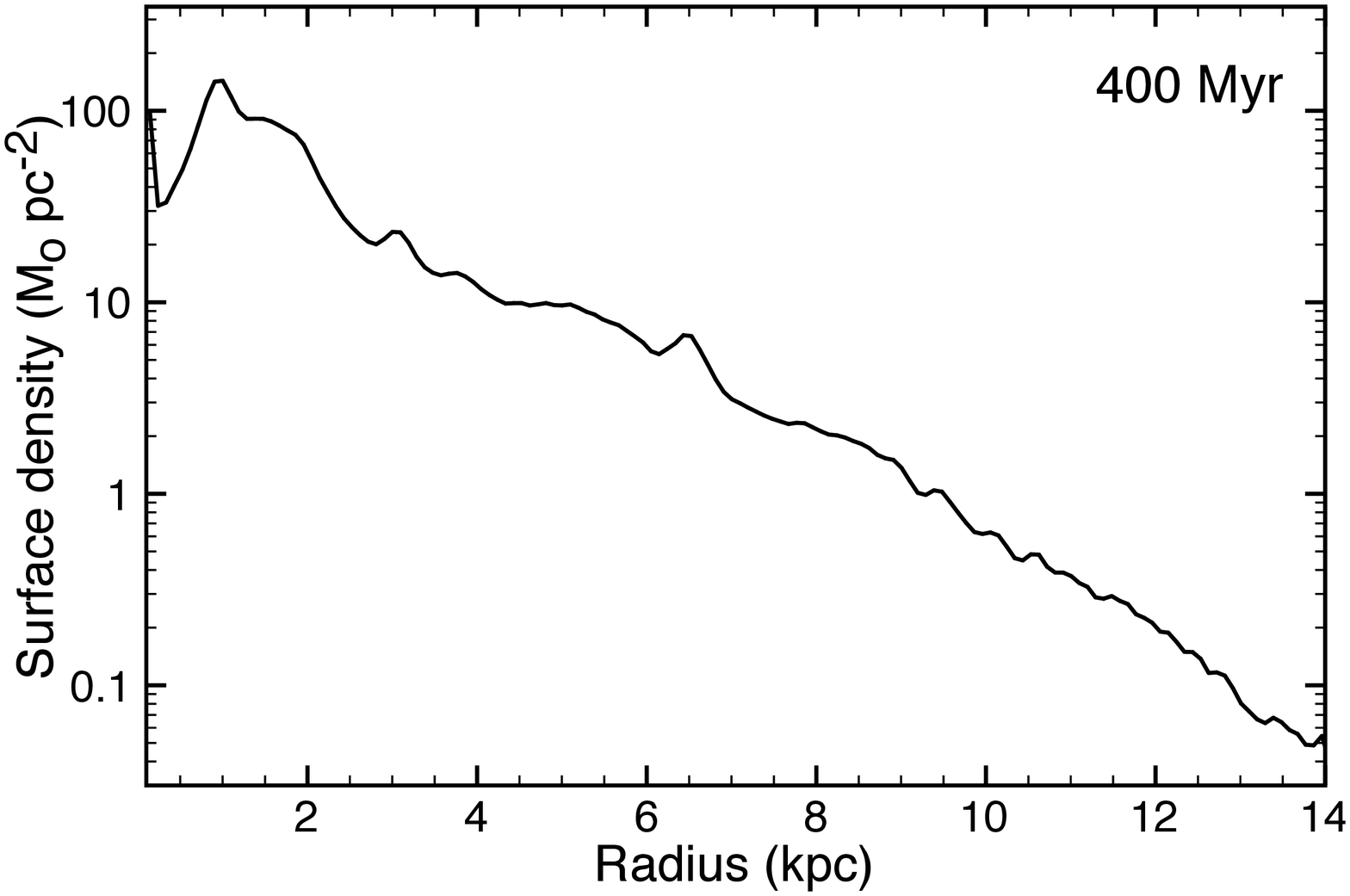}
\includegraphics[angle=0,width=2in]{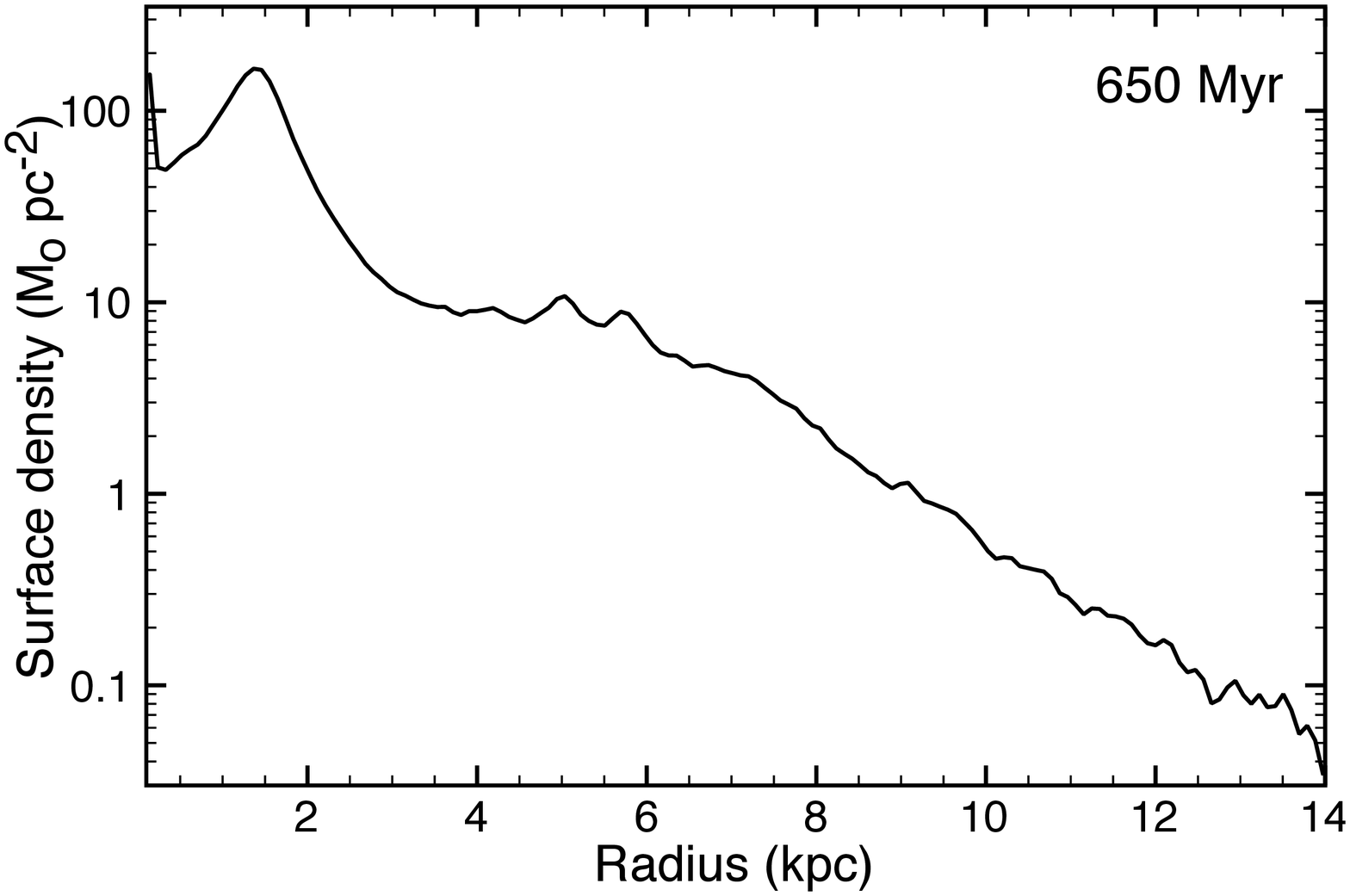}\\
\includegraphics[angle=0,width=2in]{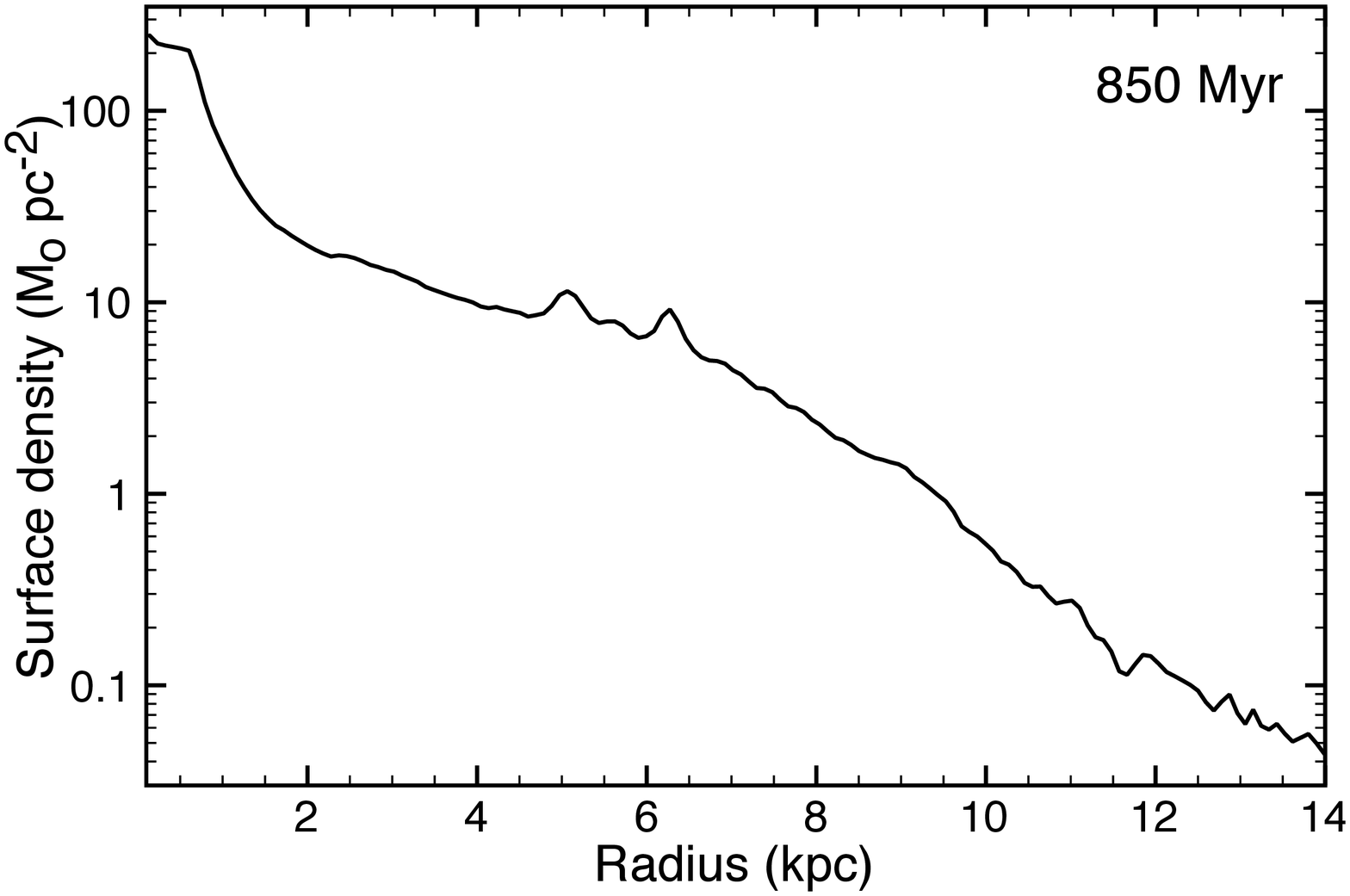}
\includegraphics[angle=0,width=2in]{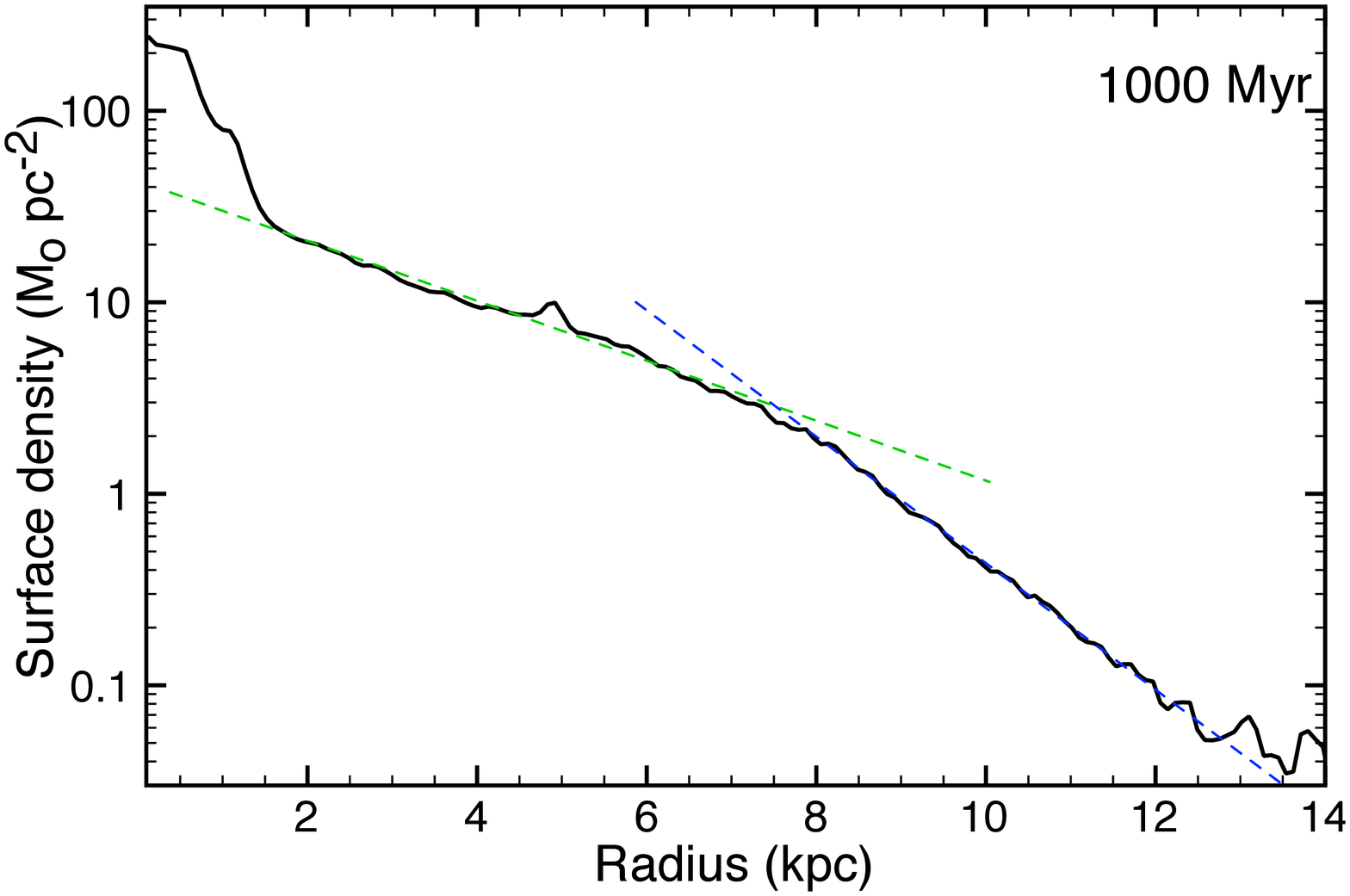}
\caption{Radial density profiles of the stellar disk in run~0, corresponding to each snapshot on Figure~\ref{fig:evol}. The evolution through a clump cluster phase changes the luminosity profile from irregular to exponential with a central bulge; an outer double-exponential component is also formed. The fits of the two exponential components in the 1--7 and 7--13~kpc radial ranges is shown on the last panel.}\label{fig:radial}\end{figure}

Figure~\ref{fig:radial} shows the azimuthally averaged radial
profiles of the disk in the fiducial run corresponding to each
snapshot in Figure~\ref{fig:evol}.
The disk starts with a flat profile that quickly becomes irregular
during the clump cluster phase. Clump evolution then forms a bulge
and an exponential disk. The exponential extends from about 1 to
7~kpc in radius with a scale length $r_{\mathrm{e}}=2.1$~kpc. All
other runs show a similar profile with a central bulge and a massive
exponential disk after 1~Gyr or less (see Table~1), while the
luminosity profile is irregular and bulgeless during the
clump-cluster phase (see also Fig.~\ref{fig:rc} for another example
in run~6).

The profile at $t=200$~Myr in Figure \ref{fig:radial} and the
corresponding image in Figure \ref{fig:evol} resemble the
azimuthally averaged profile and image of UDF1666 (Fig.
\ref{fig:udf1666}), suggesting that this clump cluster galaxy 
is already in a somewhat advanced stage. It has no bulge yet 
but a nearly-central clump that could be a proto-bulge, and has begun 
to acquire a radial density gradient, but not yet a regular exponential. 
Some other clump clusters do not have any central clump and even 
more irregular profiles (like UDF~82+99+103 and UDF~3752 in EEVFF05) 
and these ones correspond to earlier stages of the clump-cluster evolution 
according to our models.

In our models with various initial conditions (see Table~1), the clumps 
gather $f_\mathrm{C}$=23~to~38\% (average for all runs: 33\%) 
of the system baryonic mass at the peak of the clump-cluster
phase. This fraction decreases with time because shear and tidal
fields distort the clumps, and they form stars which makes them
easier to disrupt. Hence, all the mass from the clumps does not
reach the bulge: $f_\mathrm{B,C}$=35--62 (average 50)\% of the clump
mass content is finally found in the bulge. The other half of the
clump mass has been released in the disk. This way, the bulge gets
most of its mass ($f_\mathrm{C/B}$=65--74, average 69\%) from the
clumps, the rest of its mass coming from material initially present
at the disk center and severely heated and thickened when the clumps
reach the inner regions and merge together.

As for the disk component, 17--26\% of its mass
(average $f_\mathrm{C/D}$=23\%) is made of material released
by the clumps during their migration towards the disk center.
Hence the clumps have two effects on the disk:\\
{\it (i)} they release mass into the disk, which has been driven
inward by the clump migration, making the disk density
profile more concentrated\\
{\it (ii)} they act on the disk material that is not in the clumps, and
redistribute it through gravity torques, in the same manner as
spiral density waves (e.g. Pfenniger \& Friedli 1991), but more
rapidly (see Section~4.1).

Hence, the material released by the clumps and the material from the
rest of the disk both evolve into a more concentrated distribution,
which our simulation shows is an exponential disk with a bulge. We
suggest later (Section~\ref{cc-spiral}) that it is almost entirely
the clumps that are responsible for this evolution.

In addition to the main exponential disk, most runs actually end up
with a second, outer exponential profile. In the fiducial run, the
second exponential is from 7 to 13~kpc with a scale length of
$r'_{\mathrm{e}}=1.6$~kpc (Fig.~\ref{fig:radial}). At least six of
the eight runs show such a double-exponential, all with a change in
scalelength at around 7~kpc, comparable to the initial disk radius.
It is noticeable that the knee in the luminosity profile occurs at
the outer edge of the region initially affected by clump
instabilities. One can see in Figure~\ref{fig:evol} that the clumps
fling out some material in the form of spiral arms emerging from
them: this feeds the outer disk component, which has a lower density
and a steeper profile than the main exponential disk where the
clumps evolve. The scale lengths $r_{\mathrm{e}}$ and
$r'_{\mathrm{e}}$ indicated in Table~1 correspond to the 1--7~kpc
and 7--13~kpc ranges\footnote{The scale lengths have been computed
as the best fitting parameter over the 2--6 and 8--12~kpc range, but
are generally representative of the 1--7 and 7--13~kpc regions, as
seen on Figure~\ref{fig:radial}}. Runs 5 and 6 have a change of
slope that is small and might not desserve to be considered
double-exponential: the double-exponential fit is robust in the
models, but the change of slope could be difficult to detect in real
images with a limited sensitivity. In the six other runs the ratio
of the inner to outer exponential slopes is on average $\simeq 1.7$.
Double-exponentials have been observed in local (e.g. Pohlen et al.
2002) and $z=0.6-1$ spirals (P\'erez 2004), as well as in local
irregulars (Hunter \& Elmegreen 2006). The ratio of slopes for the
observations at $z\simeq 1$ (P\'erez 2004) is $2 \pm 0.35$,
compatible with those found in the present models.

\section{Discussion}

\subsection{The role of clumps and other processes in making
exponential disks}\label{cc-spiral}

We find that massive clumps in a primordial disk galaxy form an
exponential stellar disk with a bulge. There are other
theoretical ways to make
exponential disks too (e.g., Freeman 1970, Fall \& Efstathiou 1980,
Robertson et al. 2004). Pfenniger \& Friedli (1991) suggest that
spiral arm torques could make exponential profiles, although they
started with a concentrated, near-exponential profile anyway (here
we started with a flat profile).

To understand the role of clumps and spirals, we compared our
clump-free control run~C-4 with our clump-cluster model run~4. C-4
develops spiral arms; after 1~Gyr its luminosity profile is somewhat
concentrated but still irregular and far from
exponential (Fig.~\ref{fig:control}). This indicates that:\\
{\it (i)} the formation of the exponential disk in evolving clump
cluster and chain galaxies is caused by the clumps, i.e., by
the mass they release and by their gravitational action on the rest of the disk.\\
{\it (i)} the timescale to redistribute mass into an exponential
disk through spiral arms is significantly larger than the timescale
to redistribute mass by clump interactions. For example, the
exponential profile in the model by Pfenniger \& Friedli (1991) took
5~Gyr to form. Massive disk clumps are therefore expected to
dominate the formation of exponential disks.

Furthermore, if the formation of exponential-like profiles were
driven only by density waves, then there should be non-clumpy spiral
disks with non-exponential profiles that have not yet had time to
form exponentials. This contradicts the observation that irregular
profiles are highly correlated with clump cluster and chain galaxies
(EEVFF05). Also, the spiral mechanism would not work for dwarf
irregulars, which have exponential disks too (Hunter \& Elmegreen
2006). More likely, spirals adjust, maintain, and lengthen the first
exponential as random accretions and interactions distort the disk
over time. For example, spiral arms could increase the scale length
over time in the period from $z=1$ to $z=0$ (EEVFF05). Disk shear
viscosity with star formation can also make exponentials (Lin \&
Pringle 1987) but the timescale is too long, dwarfs with little
shear have exponentials, and exponentials in the far outer regions
could not be made this way if star formation is absent.

\subsection{Modelling of the interstellar medium}

The interstellar medium has been modeled with sticky particles. Such
schemes do not strictly conserve angular momentum, but the actual
loss is always much smaller than the angular momentum transferred by
spiral torques in regular spirals (Bournaud, Combes \& Semelin 2005). It should be
negligible compared to angular momentum redistribution by clump
interactions in clump clusters. To check this we ran simulation SP-0 conserving
angular momentum during inelastic collisions between gas particles
(described in Section~2.2). In this run, the bulge-to-disk mass
ratio was slightly smaller, and the clumps released a slightly
higher fraction of their mass in the disk, indicating that they are
less gravitationally bound (Table~1). The changes are minor,
however, and smaller than those corresponding to variations in
$\sigma$ or $H/D$.

Feedback from star formation has not been included because its
effects are not well known and it should not be excessively
important. For example, feedback could heat the gas in the clumps
and cause them to release mass more rapidly. The clumps 
already have masses of a few $10^8$ to $10^9$~M$_\sun$,
which is an order of magnitude above proposed masses of objects that
could be severely affected by supernovae (Dekel \& Woo 2003). The
effect of gas release would be to increase the disk mass relative to
the bulge, which would strengthen our conclusion that clump clusters
can evolve into spiral-like galaxies with massive disks. Variations
in the gas fraction for the initial disk and for the clumps does not
strongly affect the results (see Table~1, runs 4 and 5 vs. run 0).
Clumps with higher gas fractions are more tightly bound because the
mass in them is more concentrated, and then a slightly higher
fraction of the clump mass reaches the bulge.

\subsection{Comparison to real clump cluster and chain galaxies,
and the thickness of disks}

\begin{figure}
\centering
\includegraphics[width=3in]{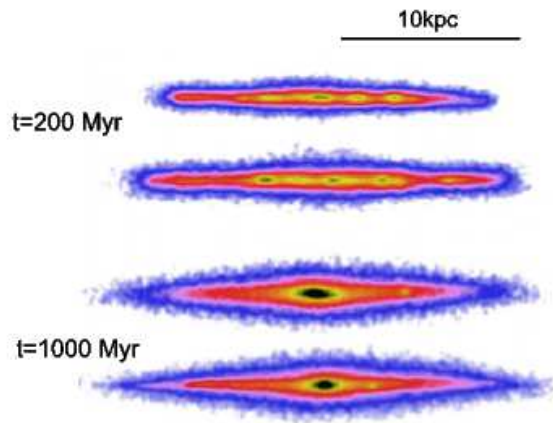}
\caption{Edge-on counterparts of the snapshots shown on Figure~\ref{fig:evol} at two instants, for two perpendicular line-of-sights each. When observed edge-on, clump clusters take the appearance of chain galaxies, with several massive clumps aligned within a bulgeless underlying disk.}\label{fig:edge}\end{figure}

Figure~\ref{fig:edge} shows an edge-on view of the fiducial model of
Figure~\ref{fig:evol}. It resembles a chain galaxy with clumps aligned
along the axis of a thick disk and no central bulge. This supports
the statistical inference that observed chains are edge-on clump
clusters.

The lifetime of the clump cluster phase in our model ($\sim0.5$~
Gyr) is consistent with the average age of $\sim300$~Myr observed
for clumps (EE05). Our simulations suggest that disk galaxies
undergo only one clump-cluster phase. After this, the resulting
bulge and exponential disk cause the galaxy to be classified as a
spiral. Large clumps can still form in the spiral disk, but the
system is more stable at this stage because of the lower gas mass
and higher velocity dispersion, so the clumps will be smaller and
less important dynamically. Multiple clump-cluster phases could be
possible only if a significant mass of gas accretes relatively quickly
onto the disk after the first phase, making it unstable to form
giant clumps again. Presumably these clumps would also disperse and
add stellar mass to the disk and bulge.

\begin{figure}
\centering
\includegraphics[angle=0,width=4in]{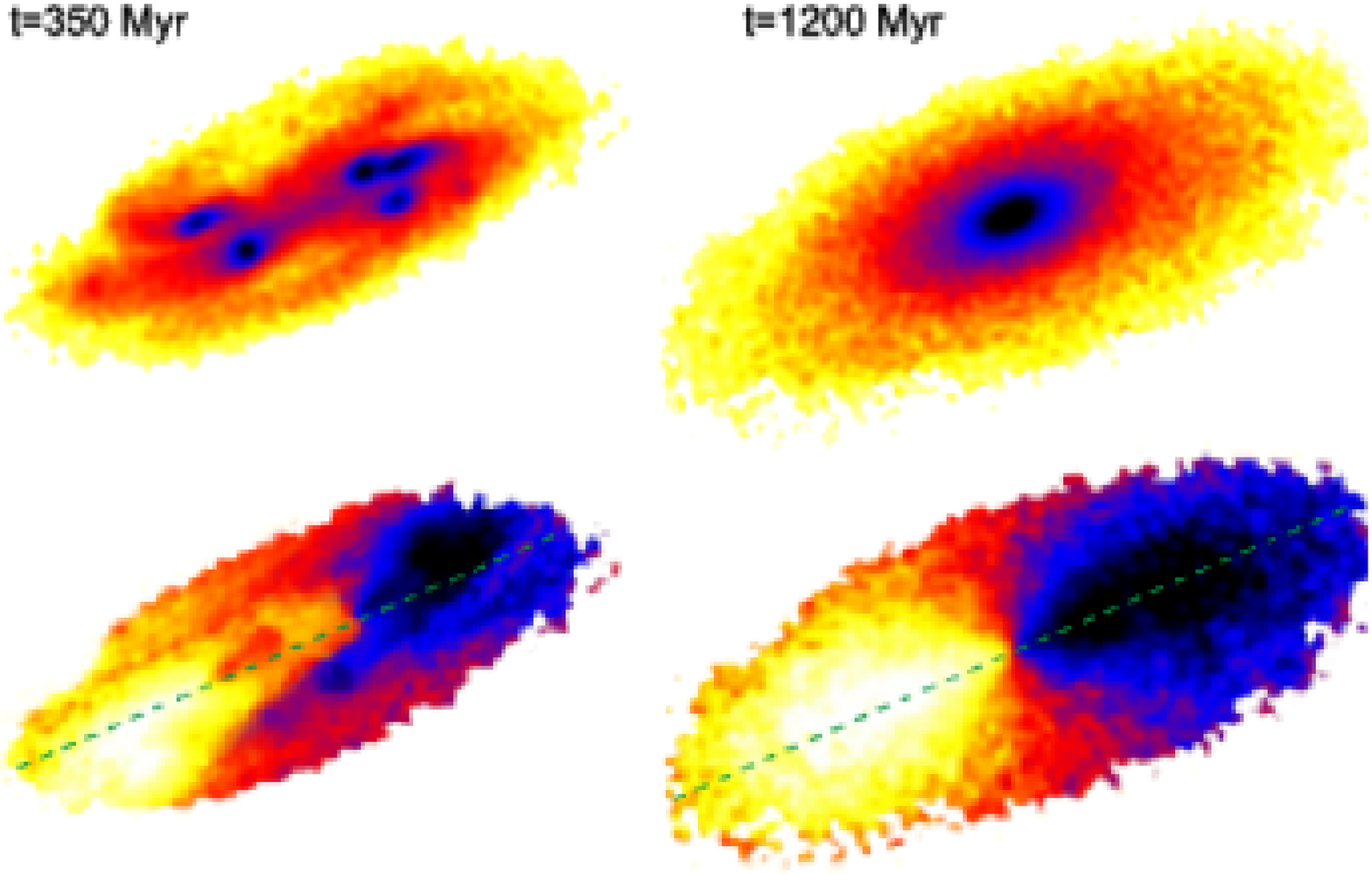}\\
\includegraphics[angle=0,width=2in]{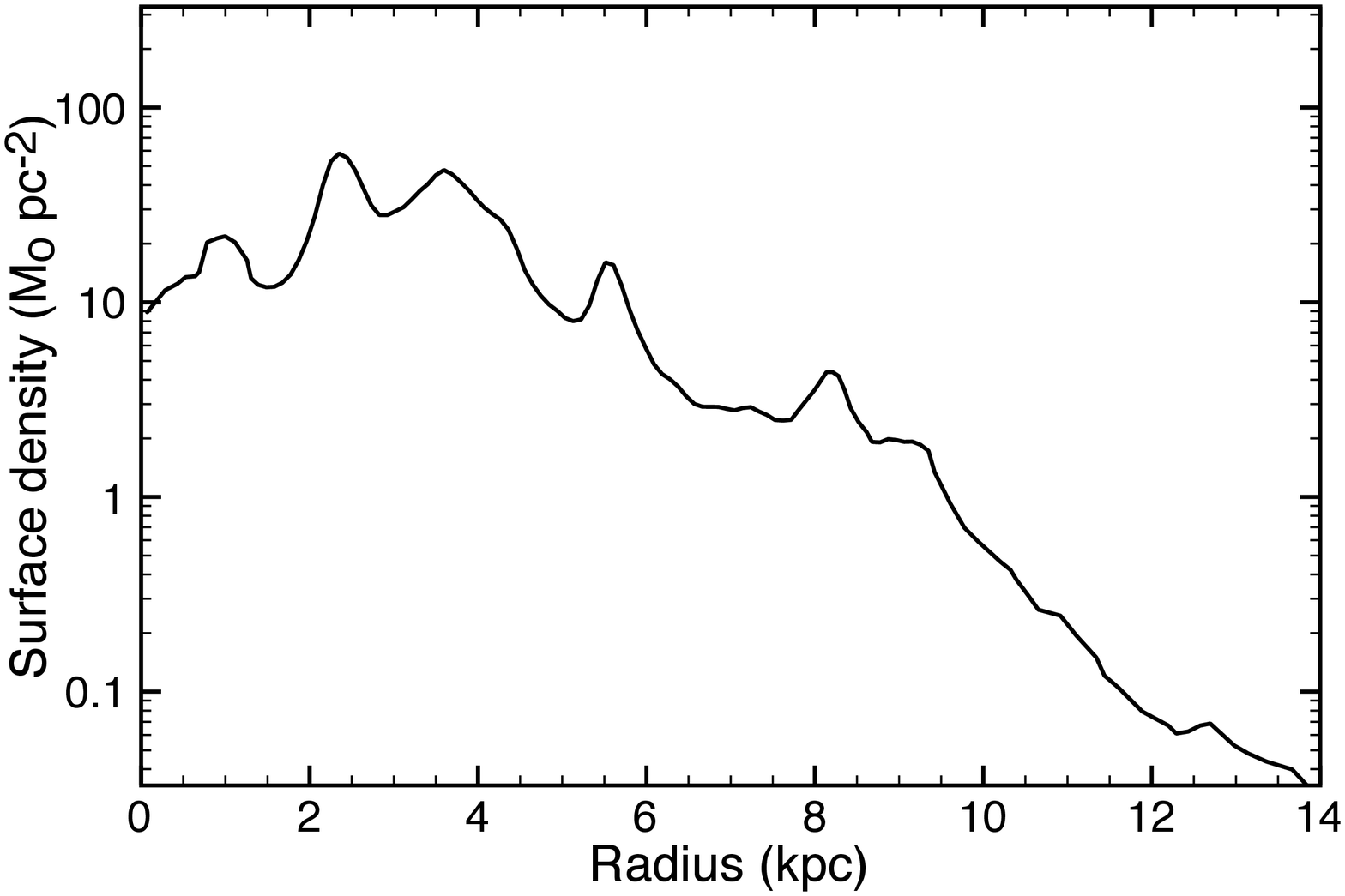}
\includegraphics[angle=0,width=2in]{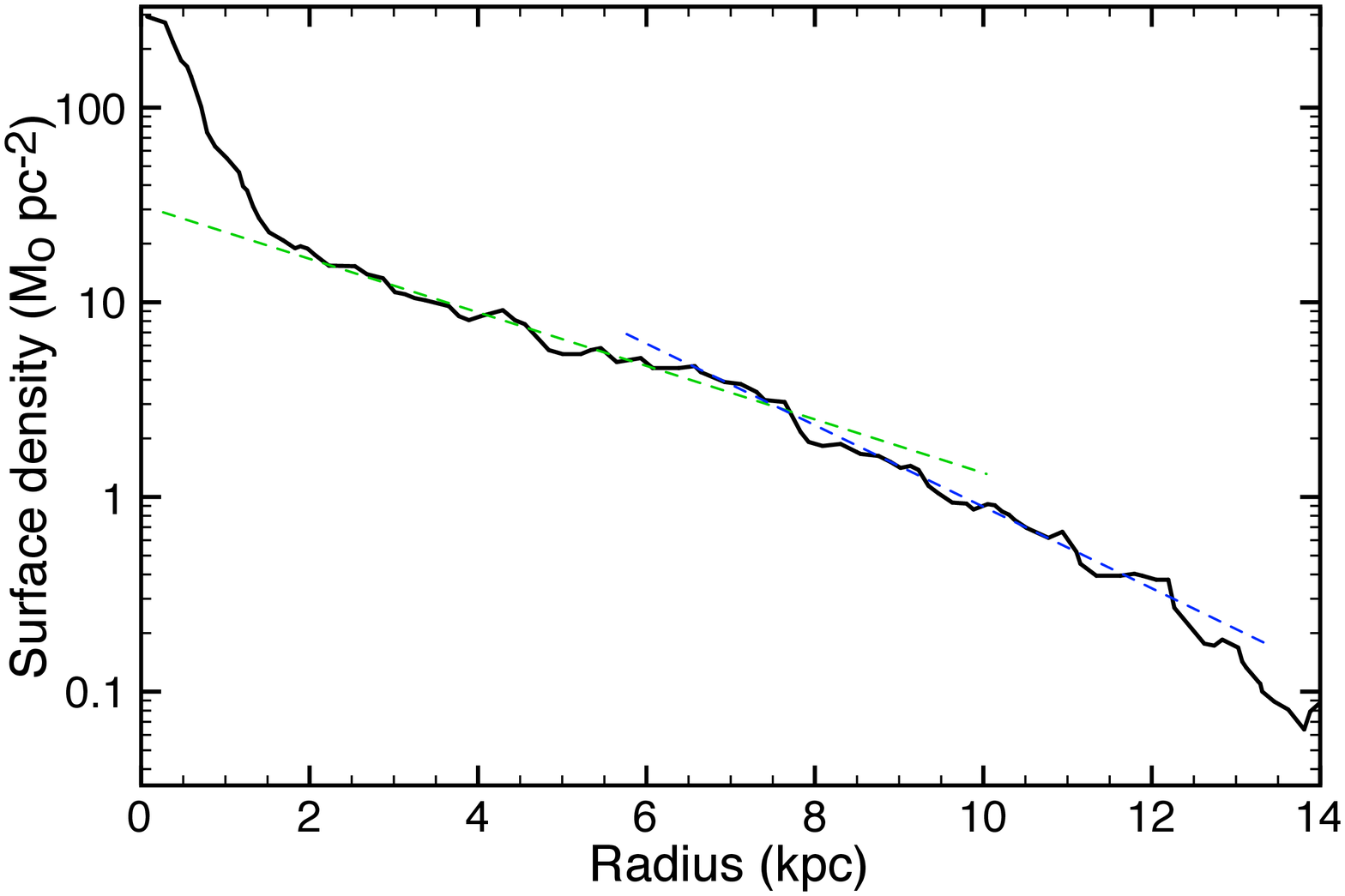}\\
\includegraphics[angle=0,width=2in]{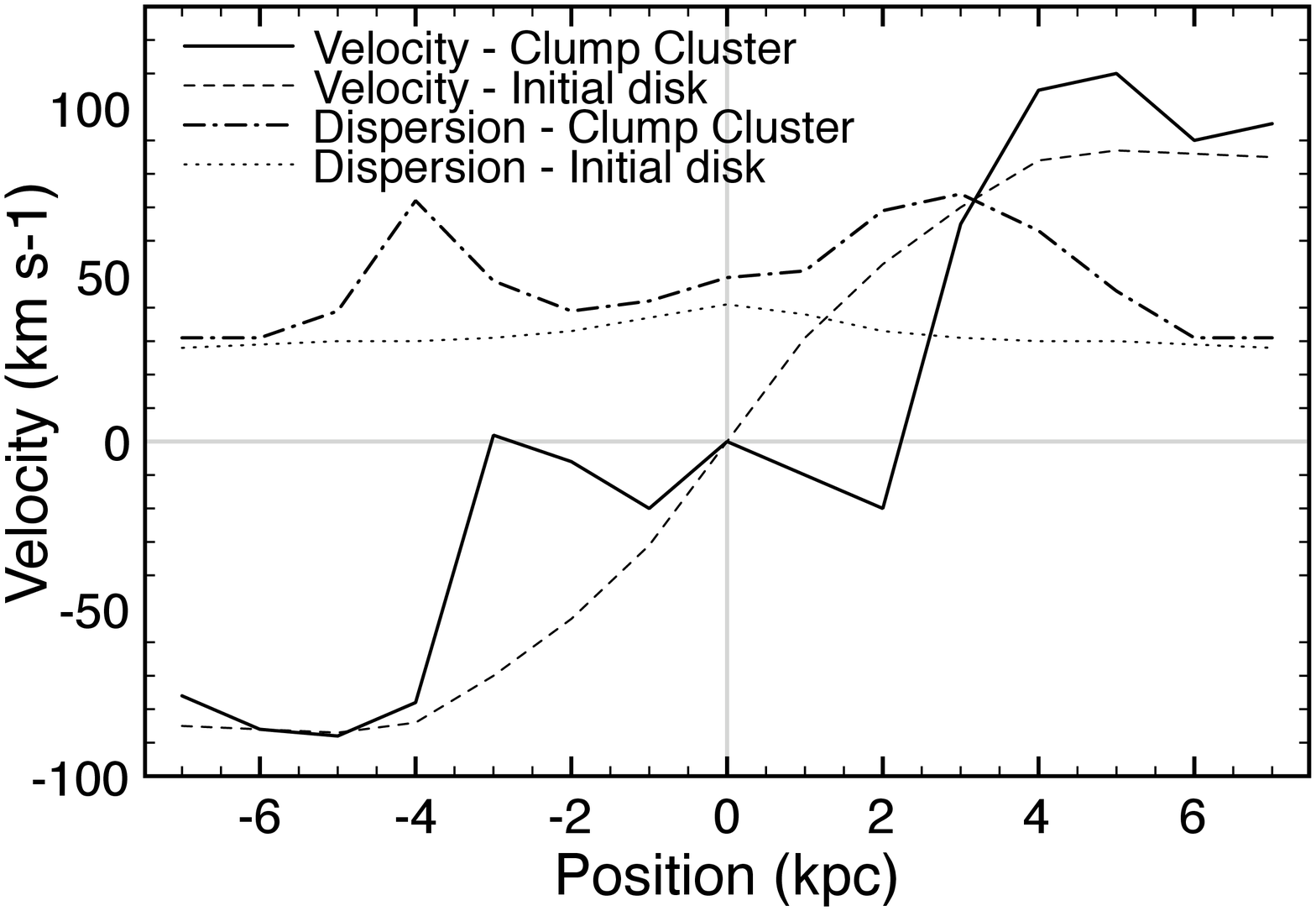}
\includegraphics[angle=0,width=2in]{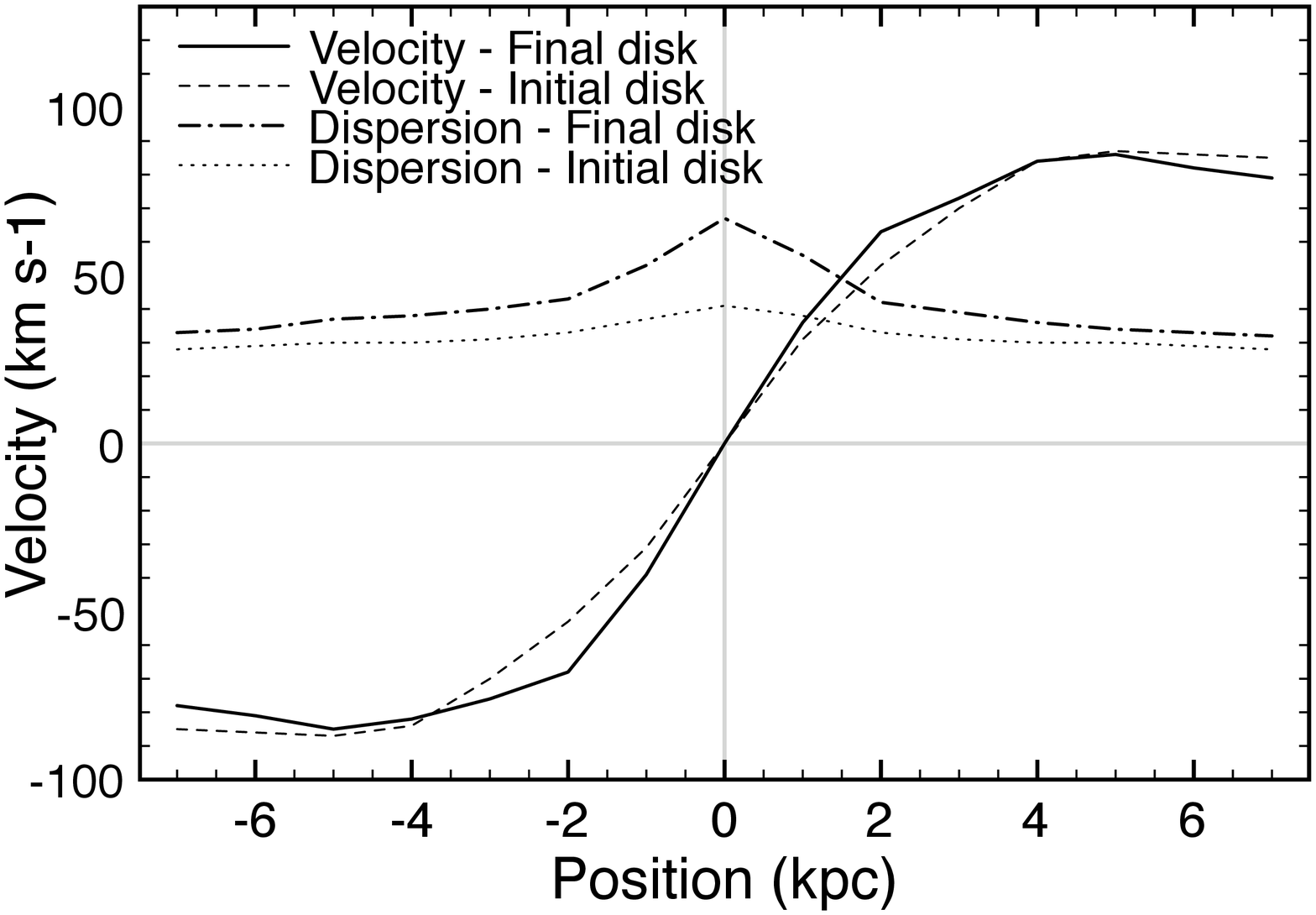}
\caption{Stellar surface density, velocity field, rotation and dispersion curves, for run~6 at two different instants (during the clump cluster phase and in the final spiral galaxy). The radial density profile are also shown. The model is viewed with 70 degrees inclination, and the velocity curves are measured along the major axis indicated by the dashed line on the velocity field. The velocities here are not deprojected ; deprojected values of $V/\sigma$ are given for all models in Table~1. Clump clusters are rotating systems, but their velocity field is highly disturbed and rotation could be difficult to detect.}\label{fig:rc}\end{figure}

Figure~\ref{fig:rc} shows the velocity fields and rotation curves
for Run~6 at each time step, obtained by viewing the disk at a 70
degree angle. During the clump-cluster phase, the linewidth is
broadened and the rotation pattern is irregular because of
clump-clump interactions. In the final galaxy, the velocity field is
typical for spirals, although the dispersion is somewhat high. We
predict that rotation in clump-clusters should be difficult to
observe because of the turbulent and irregular motions. For example,
if the outer parts of the disk in Figure~\ref{fig:rc} were not
detected beyond the clumpy region, no organized rotation would be
seen at all; only the high velocity dispersion would be evident.
This may explain why Weiner et al. (2006) found that only 30\% of
the chain galaxies in their survey had clear evidence for rotation.
F{\"o}rster Schreiber et al. (2006) and Genzel et al. (2006) also
found velocity dispersions higher than usual for clumpy,
high-redshift galaxies.

In our models, the average values of deprojected $V/\sigma$ measured
between 3 and 6~kpc in radius at the end of the simulations are
given in Table~1. They average to 2.7, which corresponds to a high
velocity dispersion but still dominated by rotation. Presumably the
dispersion of gas and new stars during the subsequent evolution will
be much lower because there are no giant clumps to stir up the disk.
Then the final stage will be a relatively quiet and thin rotating
disk with a bulge and hot disk component from the clump cluster
phase.

The final edge-on disk (Fig.~\ref{fig:edge}) appears to be about the
same thickness as the chains and edge-on spirals in the UDF, namely
$\sim1$ kpc (Elmegreen \& Elmegreen 2006b). Actually, the population
of stars already existing at the beginning of the simulation has a
scale-height of 1.3~kpc, having been heated by the violent
instabilities. The stars formed during the simulation have a
scale-height of $h=650$~kpc (defined as in Section~2.2). Hence the
final system is not made up only of a thick disk but has a massive
thick component in addition to a younger and thinner stellar disk.
The thick and thin disk components obtained here have comparable
masses, which is compatible with observations of same-mass spirals
($V_\mathrm{circ}\simeq 120$~km~s$^{-1}$ and below, Yoachim \&
Dalcanton 2006). This is also compatible with the disks at high
redshift being thick or having a massive thick component that gets
redder with height (Elmegreen \& Elmegreen 2006b). Higher-mass
present-day spirals like the Milky Way have higher thin-to-thick
disk mass fractions (see Robin et al. 2003 for the Milky Way),
suggesting that they further evolved after the primordial unstable
phases, possibly having a massive thin disk grown by gas accretion
(Bournaud \& Combes 2002, Dekel \& Birnboim 2006). The relatively
thicker disks at high redshift compared to local spirals
(Reshetnikov, Dettmar \& Combes 2003) also suggests that present-day
large spirals kept on evolving through a different process after the
clump-cluster phase. Local spirals also have larger disks (Trujillo
\& Pohlen 2005) with larger exponential scale lengths (EEVFF05),
which further indicate that the clump-clusters and chains at high
redshift formed the first exponential disks, and afterwards secular
evolution and mass accretion enlarged them to form the present-day
spirals.

\subsection{Nuclear Black Hole formation}

According to Kormendy \& Richstone (1995) and others, bulges always
have black holes (BHs) with a mass proportional to the bulge mass.
This suggests that any bulge formation model should also account for
the simultaneous formation of central BHs. There should not be
independent BH and bulge formation mechanisms. In this case, our
model requires that intermediate mass BHs (IMBHs) form in the giant
clusters of clump-cluster galaxies and migrate inward with the
cluster cores. They would then have to coalesce in the galactic
nuclei to make supermassive BHs. Simultaneous gas accretion could
contribute too. To form QSOs at high redshift, this process would
have to occur very quickly. While this seems possible for the
densest and earliest disks that form -- it takes only several orbit
times for the giant clumps to make a bulge -- cosmological and
detailed disk simulations will be necessary to confirm the basic
model. One prediction would be that clumps in clump cluster and
chain galaxies should contain IMBHs, which implies the clumps should
be X-ray and radio sources if self-absorption is not too large.
Present-day observations cannot resolve the clumps well enough to
determine if this is the case. An alternative model suggests that BHs 
form directly from accreted primordial gas without passing through 
stars first (e.g., Di~Matteo et al. 2007). Observations of IMBHs in 
disks may be able to distinguish between these two possibilities.

The clump-cluster scenario for nuclear BH formation is consistent with 
one of the prominent models in the literature. Ebisuzaki et al. (2001) showed
with N-body simulations that ultra-dense clusters can form BHs from
stellar coalescence. They proposed that nuclear BHs grow from the
coalescence of these cluster BHs and obtained the observed BH/bulge
mass ratio of $10^{-3}$. Portegies~Zwart \& McMillan (2002) found a
similar result;  massive clusters with relaxation times less than 25
Myr have mass segregation, rapid core collapse, and runaway stellar
coalescence that forms supermassive stars up to $10^{-3}$ of the
cluster mass. Portegies~Zwart et al. (2004) applied this model to
explain a suspected IMBH in a supermassive cluster of M82. G\"urkan,
Freitag \& Rasio (2004) studied a cluster with $10^7$ stars; when
the core-collapse time is less than the O-star lifetime, runaway
collisions form IMBHs with masses that are always about $10^{-3}$ of
the cluster mass.  More detailed models of stellar coalescence in
dense cluster cores confirmed the formation of supermassive stars
(Freitag, G\"urkan, \& Rasio 2006).

Portegies~Zwart et al. (2006) followed the evolution of dense
clusters in the Milky Way nuclear region and found that 10\% form
IMBHs during their inward migration. These IMBHs coalesce with the
nuclear BH at a high enough rate to account for all of the nuclear
BH mass. Matsubayashi, Makino, \& Ebisuzaki (2007) also studied the
coalescence of IMBHs to make supermassive BHs in galactic nuclei.
They found that dynamical friction becomes ineffective close to the
central BH, but that IMBH orbits become highly eccentric and spiral
in quickly anyway because of gravitational radiation.

Our model for nuclear BH formation is qualitatively similar to these
but it occurs at a much earlier stage in the life of a galaxy when
the clusters are more massive and more dispersed in the main disk.
Also in our model, the mass of the cluster BHs could be much larger
than in these other models because the early disk clusters are more
massive. If cluster BHs have $10^{-3}$ of the cluster mass, then in
our model the BHs would have $\sim10^5-10^6$ M$_\odot$.  This should
be enough to make a quick AGN in a moderate size disk galaxy, but
subsequent accretion would be necessary to make larger nuclear BHs,
and clump-cluster galaxy coalescence would be necessary to make
supermassive BHs in elliptical galaxies.

\section{Conclusions}

Our simulations suggest that gas-rich disks in the early Universe
collapse into giant star-forming clumps and resemble the clump
cluster and chain galaxies observed in the UDF. Noguchi (1999) found
similar fragmentation, as did Immeli et al. (2004b), who also
suggested such disks are a likely origin for chain galaxies. Here we
considered the observational consequences of clump interactions in
such galaxies, and we determined the resulting structures of the
galaxies.

The fragmentation of unstable primordial disks and the strong
interactions between these fragments account for the observed
kinematics of chain galaxies. Chain and clump cluster galaxies
should be found to rotate, but the heightened turbulence and
irregular orbits from clump-clump interactions make this rotation
difficult to observe. These interactions should also add a large
velocity dispersion. These predictions are consistent with the
observations of chain galaxies by Weiner et al. (2006), and with the
large velocity dispersions found in clump cluster galaxies by
F\"orster Schreiber et al. (2006) and Genzel et al. (2006).

The clumps in our simulations form very quickly. They interact with
each other, exchange angular momentum, migrate inward, dissolve, and
perturb the underlying disk. These processes form a bulge+disk
structure that is typical of spiral galaxies today, and they produce
an exponential profile for the main disk component. About half of
the clump mass goes into the disk within its initial truncation
radius, and in most simulations an outer, steeper exponential forms
from disk material that is flung out. The clump-cluster phase lasts
for $\sim$0.5--1~Gyr and occurs only once during the life of a
galaxy, unless there is a significant accretion event later.
There may be other ways to make bulges or pseudo-bulges (e.g., Fu,
Huang, \& Deng 2003; Athanassoula 2005; Kormendy \& Kennicutt 2004),
but only the present model has been observed at high redshift in
each of the important stages. Moreover, clump-cluster galaxies are
usually bulgeless (EEVFF05), so if they evolve into normal,
smooth-profile galaxies, then the bulge has to form at the same time
as the clumps disperse.  The other secular processes can only grow
and evolve this first bulge later on.

An important implication of this result is that the continuous
appearance of clump cluster galaxies throughout a wide range of
redshifts means that disk galaxies start forming over an extended
period of time. This long extent was also suggested by the star
formation timescale in comparison to the local Hubble time (EERC07).
Noeske et al. (2007) reach a similar conclusion based on the 
distribution of star formation rates over mass and time. 
When a disk becomes sufficiently massive through accretion, giant
clumps form at the local Jeans mass by gravitational instabilities.
A high velocity dispersion and a low initial disk column density put
this Jeans mass at a high value, around $10^8$ M$_\odot$. Clump
interactions redistribute disk matter into the characteristic
exponential form, and they thicken the disk to a $\sim$kpc scale
height, as observed at this stage. Clumping instabilities also cause
starbursts, with a star formation rate typically enhanced by a
factor of ten in the present models. These bursts are not related to
interactions, although the large-scale environment should influence
this kind of activity indirectly through its role in the growth of
disks. Such internal bursting could explain the existence of active
star-forming galaxies that appear to be isolated (Genzel et al.
2006). Exponential disks should continue to evolve after the
clump-cluster phase, but the evolution should be more gradual and
may involve the conventional spiral and bar torques applied to
slowly accreting disks.

Supermassive black holes correlate with bulges and have to form very
quickly in the life of a galaxy to explain QSOs and AGNs at high
redshift. Based on published models for runaway growth of
supermassive stars in dense cluster cores, we speculate that the
clumps in clump cluster and chain galaxies rapidly form black holes
with masses of $10^5-10^6$ M$_\odot$, and that these black holes
migrate inward as the bulges form to make nuclear black holes. This
whole process should operate in several rotation periods of the
disk, which can be fast or slow depending on the galaxy density and
therefore on the epoch of galaxy formation. Simulations of this BH
migration process will be discussed elsewhere.

\acknowledgments

We are grateful to an anonymous referee for constructive comments that
improved the presentation of the results. The simulations in this paper
were performed on the NEC-SX8 and SX8R
vectorial computers at the CEA/CCRT and IDRIS centers.

{}

\begin{table}
\begin{center}
\caption{Run parameters and results. The various fractions are defined in Section~2.3 and the bulge-to-total mass ratio $B/T$ refers to the total baryonic mass, not counting the dark matter halo.
\label{tbl-2}}
\begin{tabular}{cccccccccccccccc}
\tableline\tableline
Run && $\sigma$ & $f_{\mathrm{G}}$ & $H/D$ && $f_\mathrm{C}$ & $f_\mathrm{g,C}$ & $f_\mathrm{B,C}$ & $f_\mathrm{C/B}$ & $f_\mathrm{C/D}$ && $B/T$ & $r_{\mathrm{e}}$ & $r'_{\mathrm{e}}$ & $V/\sigma$\\
\tableline
0 && 9   & 50\% & 0.50 &&  0.38 & 0.56 & 0.53 & 0.67 & 0.26 && 0.30 & 2.1 & 1.6 & 2.9 \\
1 && 5   & 50\% & 0.50 &&  0.32 & 0.60 & 0.48 & 0.74 & 0.22 && 0.21 & 2.9 & 1.7 & 2.7 \\
2 && 15 & 50\% & 0.50 &&  0.31 & 0.52 & 0.40 & 0.66 & 0.25 && 0.19 & 2.3 & 1.4 & 2.2 \\
3 && 20 & 50\% & 0.50 &&  0.39 & 0.47 & 0.62 & 0.74 & 0.23 && 0.33 & 2.0 & 1.6 & 2.4 \\
4 && 9   & 25\% & 0.50 &&  0.26 & 0.32 & 0.36 & 0.67 & 0.20 && 0.14 & 2.4 & 1.1 & 3.0 \\
5 && 9   & 75\% & 0.50 &&  0.34 & 0.63 & 0.61 & 0.65 & 0.19 && 0.32 & 2.0 & (1.8) & 2.8 \\
6 && 9   & 50\% & 0.25 &&  0.41 & 0.51 & 0.62 & 0.71 & 0.25 && 0.36 & 2.3 & (2.0) & 2.4\\
7 && 9   & 50\% & 0.80 &&  0.23 & 0.57 & 0.35 & 0.67 & 0.17 && 0.12 & 2.2 & 1.2 & 3.1\smallskip \\
SP-0&& 9 & 50\% & 0.50 && 0.36 & 0.55 & 0.51 & 0.68 & 0.24 && 0.27 & 2.2 & 1.4 & 3.0 \smallskip  \\
\tableline
\end{tabular}
\end{center}
\end{table}

\begin{figure}
\centering
\includegraphics[width=3in]{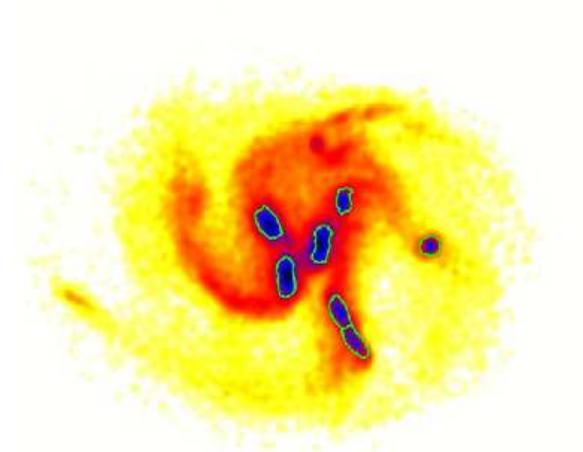}
\caption{Contour of the ``clumps'' defined as entities corresponding to overdensities (above three times the average mass density at each radius) of more than $2\times10^8$~M$_\sun$ each. Two clumps are defined in the lower right region, where the density between the two peaks is lower by factor three than each density peak. We consider only regions less extended than 3 kpc in any direction, which ensures that overdense regions in the outer parts, that are spiral arms rather than clumps, are not counted.}\label{fig:detect}\end{figure}

\end{document}